\documentclass{article}
\usepackage{spconf,amsmath,graphicx}
\usepackage{amsfonts}
\usepackage{arydshln}
\usepackage{enumitem}
\setlist{nosep, leftmargin=14pt}

\usepackage{mwe} 

\usepackage{hyperref}
\usepackage{pifont}
\title{Voxels Intersecting along Orthogonal Levels Attention U-Net for Intracerebral Haemorrhage Segmentation in Head CT}
\name{Qinghui Liu, Bradley J MacIntosh, Till Schellhorn, Karoline Skogen, Kyrre Eeg Emblem, Atle Bj{ø}rnerud\thanks{Supported and funding provided by Helse S{ø}r-{Ø}st Regional Health Authority and the Research Council of Norway under Grant 325971.}}
\address{Department of Physics and Computational Radiology, Division of Radiology and Nuclear Medicine, \\Oslo University Hospital (OUS), Rikshospitalet, 0372 Oslo, Norway}
\begin{document}
%
\maketitle
\begin{abstract}
We propose a novel and flexible attention based U-Net architecture referred to as "Voxels-Intersecting Along Orthogonal Levels Attention U-Net" (viola-Unet), for intracranial hemorrhage (ICH) segmentation task in the INSTANCE 2022 Data Challenge on non-contrast computed tomography (CT). The performance of ICH segmentation was improved by efficiently incorporating fused spatially orthogonal and cross-channel features via our proposed Viola attention plugged into the U-Net decoding branches. The viola-Unet outperformed the strong baseline nnU-Net models during both 5-fold cross validation and online validation. Our solution was the winner of the challenge validation phase in terms of all four performance metrics (i.e., DSC, HD, NSD, and RVD). The code base, pretrained weights, and docker image of the viola-Unet AI tool are publicly available at \url{https://github.com/samleoqh/Viola-Unet}.

\end{abstract}
\section{Introduction}
\label{sec:intro}
\begin{figure}[htb]
\includegraphics[width=\linewidth]{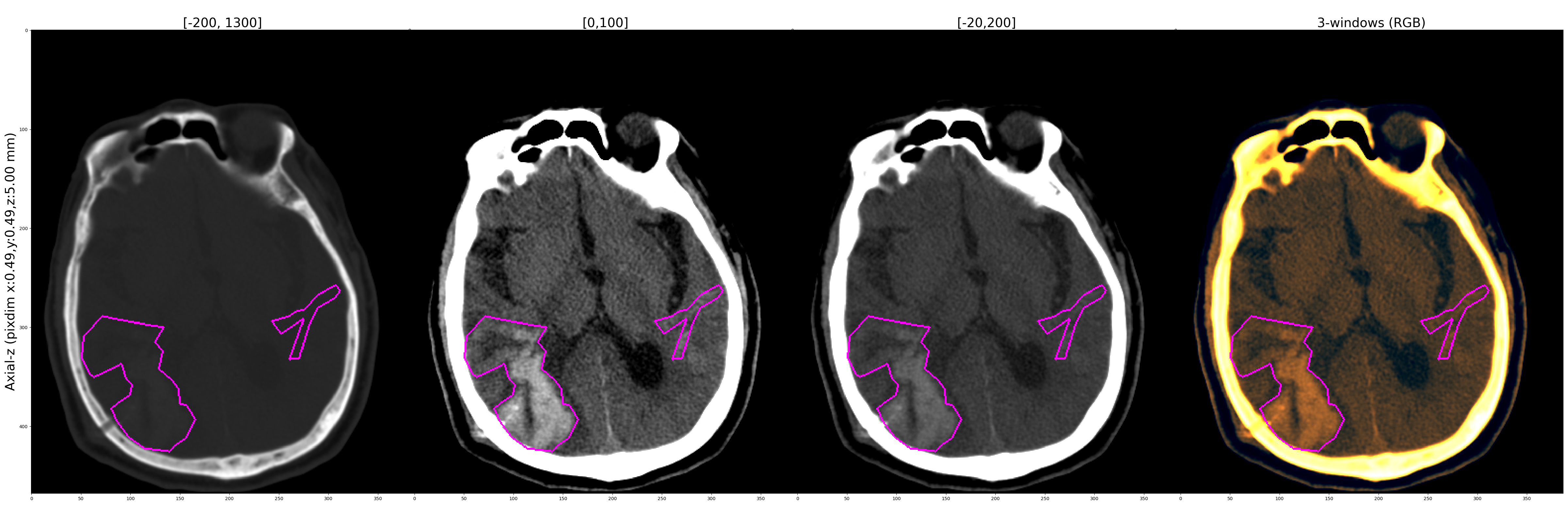}
\caption{An example from the INSTANCE2022 dataset. Each images shows different Hounsfield Unit (HU) windowing levels to take advance of an RGB-style 3-window data input combination (from left to right: $[-200 \sim 1300]$, $[0 \sim 100]$, $[-20 \sim 200]$, and $3$-window combination), with the ICH labeled regions highlighted by the pink edge lines. } \label{fig1}
\end{figure}

A spontaneous intracranial hemorrhage (ICH) is the second most common subtype of stroke and a critical disease usually leading to severe disability or death~\cite{an2017epidemiology}. Accurately estimating the volume of ICH is important in clinical diagnosis procedures for predicting hematoma progression and early mortality~\cite{broderick1993volume}. Non-contrast Computed Tomography (CT) is the most commonly used modality in regular clinical practice for diagnosing ICH. The hematoma volume can be calculated by radiologists manually with ABC/2 method~\cite{kothari-pmid8711791} in clinical practice. However, the ABC/2 method exhibits significant volume estimation error, particularly for hemorrhages with irregular shapes as shown in Fig.~\ref{fig1}. Hence, a fully automated segmentation method that allows accurate and rapid volume quantification of ICH is high desirable.

Deep learning (DL) algorithms have recently received increasing attention in computer-aided automatic methods for medical data analysis. The state-of-the-art medical image segmentation models tend to rely on the popular U-Net~\cite{ronneberger2015u} architecture, an encoder-decoder convolutional neural network (CNN) based approach with end-to-end training pipeline for pixel- or voxel-wise segmentation. Several U-Net-like models have tackled ICH segmentation using head CT scans~\cite{arab2020fast,hssayeni2020,sharrock20213d,yu-stroke2022-ichdeep} and these successes are mirrored in other brain imaging fields such as tumor segmentation of multi-modal MRI scans~\cite{futrega2021optimized,luu2021extending}. Isensee et al., in particular, used the nnU-Net framework~\cite{isensee2021nnu} to present a winning model for the BraTS20 challenge~\cite{menze2014multimodal}, with a self-configuring method for various DL-based biomedical image segmentation tasks. Thus, we chose nnU-Net as the strong baseline model in the current work.

In this paper, we propose a novel solution for the INSTANCE 2022 ICH segmentation challenge~\cite{instance2022}, with the goal of making a computationally efficient, accurate, and robust end-to-end 3D deep learning model. Our solution has won the challenge validation phase, with an average DSC score of 0.795.

\section{Methods}
\label{sec:method}
Our solution is called "viola-Unet" as it relies on Voxels in feature space that Intersect along Orthogonal Levels to provide an Attention U-Net, which is an asymmetric encoder-decoder architecture with 7-depth layers ( shown in Figure~\ref{fig2} (a)). The number of channels at each encoder was 32, 64, 96, 128, 192, 256 and 320, while the channel-numbers at each corresponding decoder layer were 32, 64, 96, \textit{128}, \textit{128} and \textit{128}. In addition, the input patch size was $3 \times 160 \times 160 \times 16$ with 2 extra scales of deep supervision outputs. 

\subsection{Viola attention module} 
\begin{figure}[htb]
\begin{minipage}[b]{1\linewidth}
  \centering
  \centerline{\includegraphics[width=8.5cm]{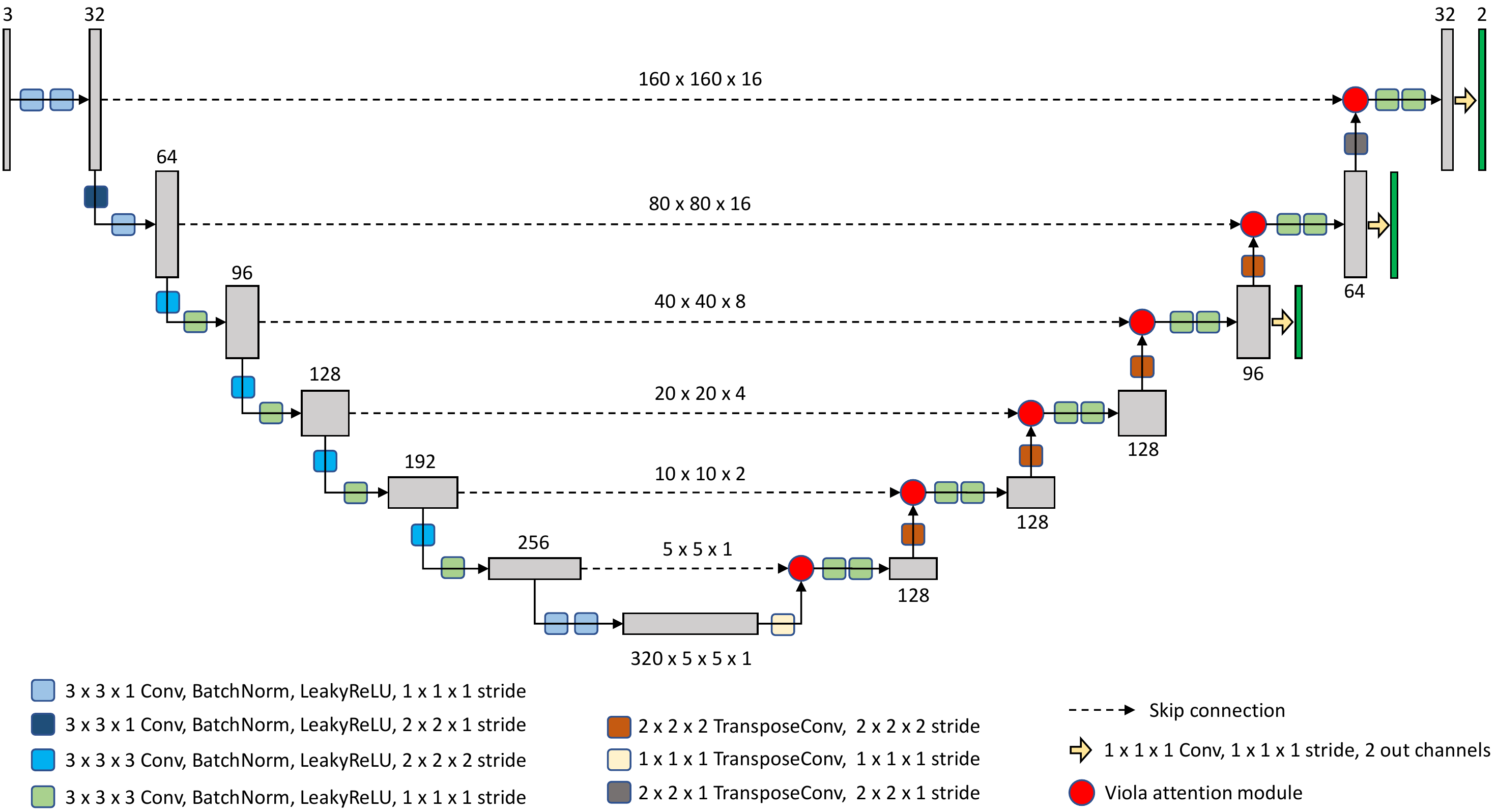}}
 \vspace{0.2cm}
  \centerline{(a) The Viola U-Net architecture.}\medskip
\end{minipage}
\begin{minipage}[b]{1\linewidth}
  \centering
  \centerline{\includegraphics[width=9.5cm]{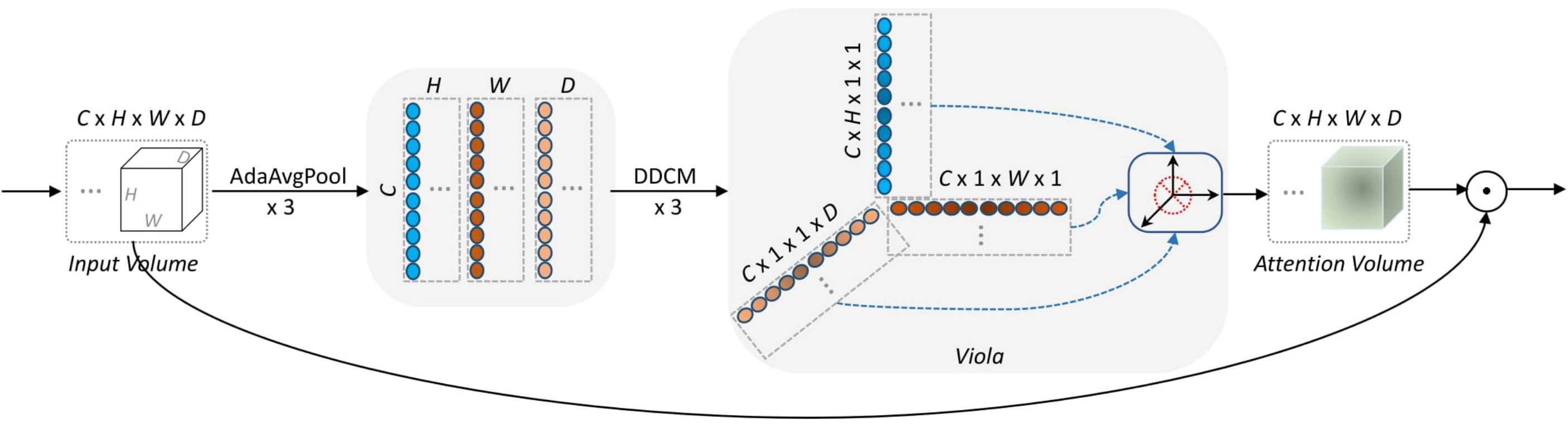}}
 \vspace{0.2cm}
  \centerline{(b) The Viola attention module pipeline. }\medskip
\end{minipage}

\caption{The Viola U-Net (viola-Unet) architecture powered by the proposed Voxels Intersecting along Orthogonal Levels Attention (viola) module. Additional two output heads are only used for deep supervision \cite{zhu2017deeply} training. Here AdaAvgPool denotes adaptive average pooling, and DDCM denotes dense dilated convolutions' merging network~\cite{liu2020TGRS}.}
\label{fig2}
\end{figure}

Squeeze-and-Excitation (SE) networks are able to recalibrate channel-wise feature responses by explicitly modeling interdependencies between channels on 2D feature planes\cite{hu2018squeeze}. The viola-Unet attention method is similar; Fig.~\ref{fig2} (b) shows how the viola attention module incorporates features along orthogonal directions, an efficient way to incorporate through-plane features. 

Overall Viola module is composed of three key blocks, i.e., the adaptive average pooling (AdaAvgPool) module that squeezes the input feature volume (e.g., $\textbf{X} \in \mathbb{R}^{C\times H\times W\times D}$, where $C, H, W$, and $D$ represent channel, height, width, and depth for a given feature volume.) into three latent representation spaces (e.g., $\textbf{X}_h \in \mathbb{R}^{C\times H} $, $\textbf{X}_w \in \mathbb{R}^{C\times W} $, and $\textbf{X}_d \in \mathbb{R}^{C\times D} $) along each axis of the input feature patch. The customized dense dilated convolutions merging (DDCM) networks~\cite{liu2020TGRS} fuses cross-channel and non-local contextual information on each orthogonal direction with adaptive kernel sizes (i.e., $k = [2(C//32) + 3, 1]$ ), dilated ratios (i.e., $dilation = [1, k, 2(k-1) + 1, 3(k-1) + 1]$ ) and strides (i.e, $strides = [(2,1), (2,1), (4,1), (4,1)]$). The Viola unit constructs the voxels intersecting along orthogonal level attention volume (i.e. $\textbf{A}_{viola} \in \mathbb{R}^{C\times H\times W\times D}$) based on fused and reshaped cross-channel-direction latent representation spaces (i.e., $\textbf{X}_h \in \mathbb{R}^{C\times H\times 1\times 1}, \textbf{X}_w \in \mathbb{R}^{C\times 1\times W\times 1}$, and $\textbf{X}_d \in \mathbb{R}^{C\times 1\times 1\times D}$), by the following mathematical equations\footnote{Unless particularly specified, we use bold capital characters for matrices and tensors, lowercase and capital characters in italics for scalars and bold italics for vectors.}.
\begin{equation}\label{eq:att1}
\begin{aligned}
 \tilde{\textbf{X}}_h, \tilde{\textbf{X}}_w, \tilde{\textbf{X}}_d &= \sigma_{s}\left(\textbf{X}_h, \textbf{X}_w, \textbf{X}_d\right),\\ 
 \hat{\textbf{X}}_h, \hat{\textbf{X}}_w, \hat{\textbf{X}}_d &= \sigma_{gt} \left(\textbf{X}_h, \textbf{X}_w, \textbf{X}_d\right)\;,
 \end{aligned}
\end{equation}

\begin{equation}\label{eq:att2}
\begin{aligned}
 {\textbf{A}} &= \sigma_{r}\left(\tilde{\textbf{X}}_h + \hat{\textbf{X}}_h +  \tilde{\textbf{X}}_w + \hat{\textbf{X}}_w + \tilde{\textbf{X}}_d + \hat{\textbf{X}}_d\right)
  \\
&\quad + \tilde{\textbf{X}}_h \otimes \tilde{\textbf{X}}_w + \tilde{\textbf{X}}_w \otimes \tilde{\textbf{X}}_d  + \tilde{\textbf{X}}_d \otimes \tilde{\textbf{X}}_h\\
&\quad + \tilde{\textbf{X}}_h \otimes \tilde{\textbf{X}}_w \otimes \tilde{\textbf{X}}_d \;,
 \end{aligned}
\end{equation}

\begin{equation}\label{eq:att3}
\begin{aligned}
    {\textbf{A}_{viola}} &= \left(\alpha +
    {\|f_{latten}\left(\textbf{A}\right)\|_{2}}^{-1}\right) \textbf{A} + \beta\;,\\
{\textbf{X}} &= \textbf{X} \odot \textbf{A}_{viola} \;.
\end{aligned}
\end{equation}
where $\sigma_{s}$ denotes the $\operatorname{Sigmoid}$ activation function, $\sigma_{gt}$ denotes a combination function of group normalization \cite{wu2018group} ($G=2$ in this work) and $\operatorname{Tanh}$ non-linearity, $\sigma_{r}$ is the $\operatorname{ReLU}$ activation operator, $\otimes$ denotes the tensor product and $\odot$ denotes the element-wise multiplication. Furthermore, two attention coefficients are introduced: $\alpha = 0.1$ to balance the weights of attention maps and $\beta = 0.3$ to weight residual feature maps.

\subsection{Architecture considerations:} The  viola-Unet is flexible and configurable, i.e. strides and kernel sizes at each layer, number of features in both encoder and decoder layers, symmetric or asymmetric, the number of deep supervision outputs. In addition, we used a self-configured U-Net architecture as a baseline model from the official open source nnU-Net framework~\footnote{\url{https://github.com/MIC-DKFZ/nnUNet}.}. The nnU-Net had a depth of 6. The number of channels at each encoder and decoder (symmetric) level were: 32, 64, 128, 256, 320 and 320. The input path size was $1 \times 320 \times 320 \times 16$ with 5 scales of deep supervision training outputs.

\section{Data, experiments and results}
\label{sec:result}
\subsection{Dataset and evaluation metrics}
The INSTANCE 2022 challenge dataset~\cite{li2021hematoma,instance2022} consists of 200 non-contrast 3D head CT scans of clinically diagnosed patients with ICH of various types, such as subdural hemorrhage (SDH), epidural hemorrhage (EDH), intraventricular hemorrhage (IVH), intraparenchymal hemorrhage (IPH), and subarachnoid hemorrhage (SAH). N=100 of the publicly available cases were used for training; the remaining N=100 cases were held-out for the validation set (N=30 for the public leaderboard, and N=70 for the competitor rankings). Model performance was evaluated by four measures: Dice Similarity Coefficient (DSC), Hausdorff distance (HD), Relative absolute Volume Difference (RVD), and the Normalized Surface Dice (NSD). 

\subsection{Implementation and training}
Our code for this study were written in PyTorch with use of the open source Monai~\footnote{\url{https://monai.io/}.} library version $0.9.0$. We adopted and modified Monai's network codes to implement the proposed models (both viola-Unet and modified nnU-Net). 

Guided by our empirical results, we trained all networks with randomly sampled patches of fixed size ($3 \times 160\times 160\times 16$) as input and a batch size of 2. Each network was trained with 5-fold cross validation for up to 72,000 steps using stochastic gradient descent (SGD) and an optimizer with Nesterov momentum of 0.99. The initial learning rate was $7 \times 10^{-3}$ with applying a cosine annealing scheduler~\cite{loshchilov2016sgdr} to reduce the learning rate over epochs. We used a linear warm-up learning rate during the first 1000 steps. A sliding window inference method was applied to evaluate the model on the local validation set after every 200 training steps. We stored the checkpoint with the highest mean dice score on the validation set of the current fold during the training phase. Based on our training observations to achieve fast and stable convergence for each network, we applied a combination loss function of the dice loss~\cite{milletari2016v} and Focal loss~\cite{lin2017focal} for all our experiments.

\subsubsection{self-training strategy}
We also utilised self-training strategy to do semi-supervised fine-tune learning on online validation dataset. The semi-supervised learning principle with self-training algorithms is to train a model iteratively by assigning pseudo-labels to the set of unlabeled training samples in conjunction with the labeled training set \cite{amini2022self}. In practice, we manually select the best prediction on each validation example from each submission as the pseudo-label and put them into our training set to fine-tune our models repeatedly.

\subsection{Results}
\begin{figure}[htb]
\centering
\includegraphics[width=\linewidth]{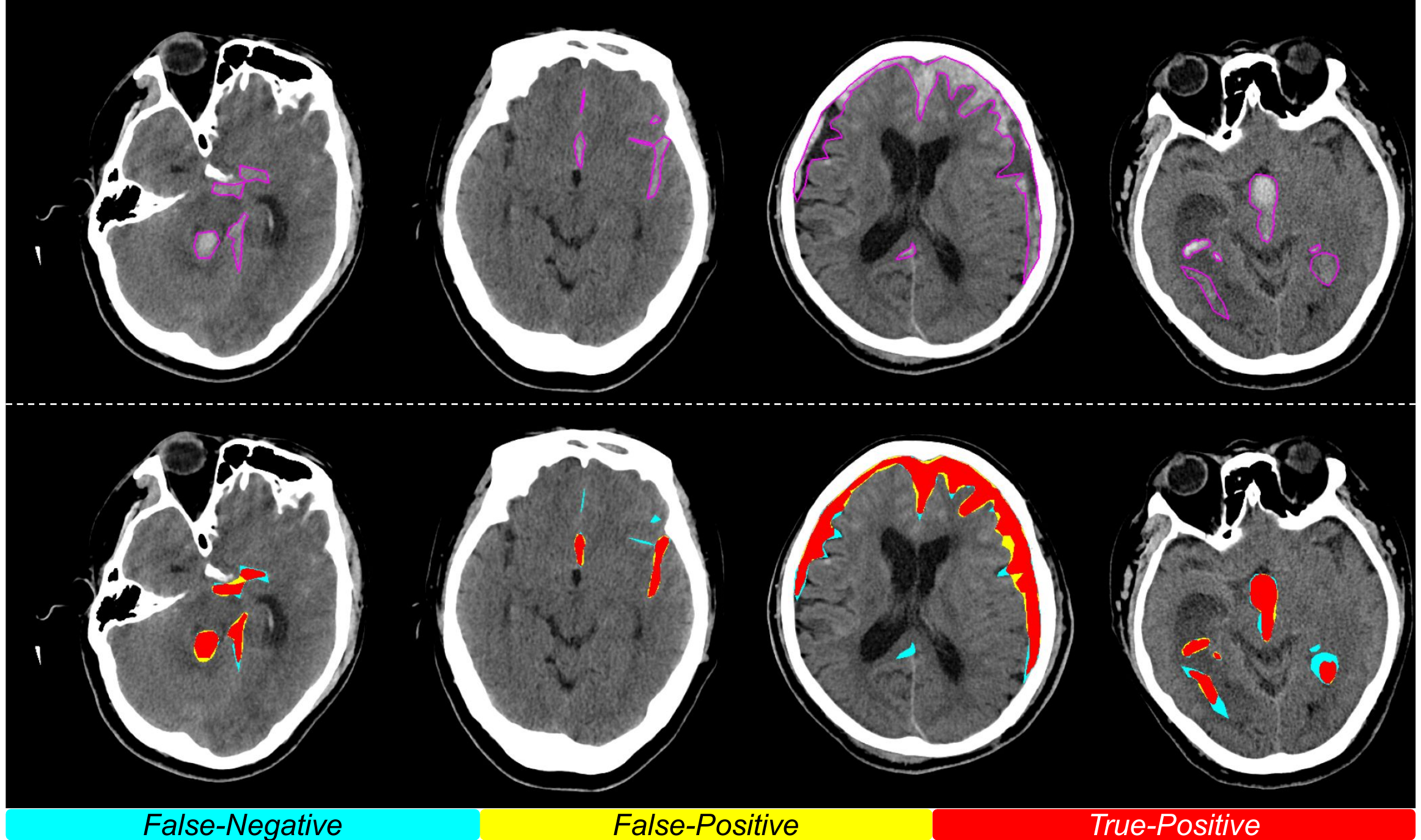}
\caption[Axial view segmentation results]{Axial view segmentation results with the Viola-Unet model on local cross-validation sets. The top row represents the input images with ground-truth regions highlighted with pink edge lines, and the bottom row represents the model segmented results, where areas colored in red denote a true positive (TP), yellow a false positive (FP), and light blue a false negative (FN).} \label{fig:fig3}
\end{figure}

Table \ref{tab1} shows the average DSC scores for each 5-folds with the nnU-Net baseline models and viola-Unet models, respectively. The viola-Unet outperforms the baseline nnU-Net by a significant margin (mean DSC $+2.18\%$). Fig.~\ref{fig:fig3} shows some examples of the predictions with the Viola-Unet model on local cross-validation sets. 

\begin{table}
\begin{center}
\caption{Average DSC for each of the 5-folds. Results for a base nnU-Net configuration are shown along with a smaller-sized version of viola-Unet (s denotes small).}\label{tab1}
\resizebox{.4\textwidth}{!}{
\begin{tabular}{|c|c|c|}
\hline
\textbf{Model} & \textbf{nnU-Net-base} & \textbf{viola-Unet-s}\\
\hline
Fold 0  & 0.7562  & \textbf{0.7786}\\
Fold 1  & 0.7345  & \textbf{0.7530}\\
Fold 2  & 0.7796  & \textbf{0.7990}\\
Fold 3  & 0.7555  & \textbf{0.8058}\\
Fold 4  & \textbf{0.7746}  & 0.7730\\ \hline
Mean DSC  & 0.7601  & \textbf{0.7819} ($+2.18\%$)\\
\hline
\end{tabular}}
\end{center}
\end{table}

\begin{figure*}[htb]
\begin{minipage}[b]{0.5\textwidth}
  \centering
  \centerline{\includegraphics[width=7.5cm]{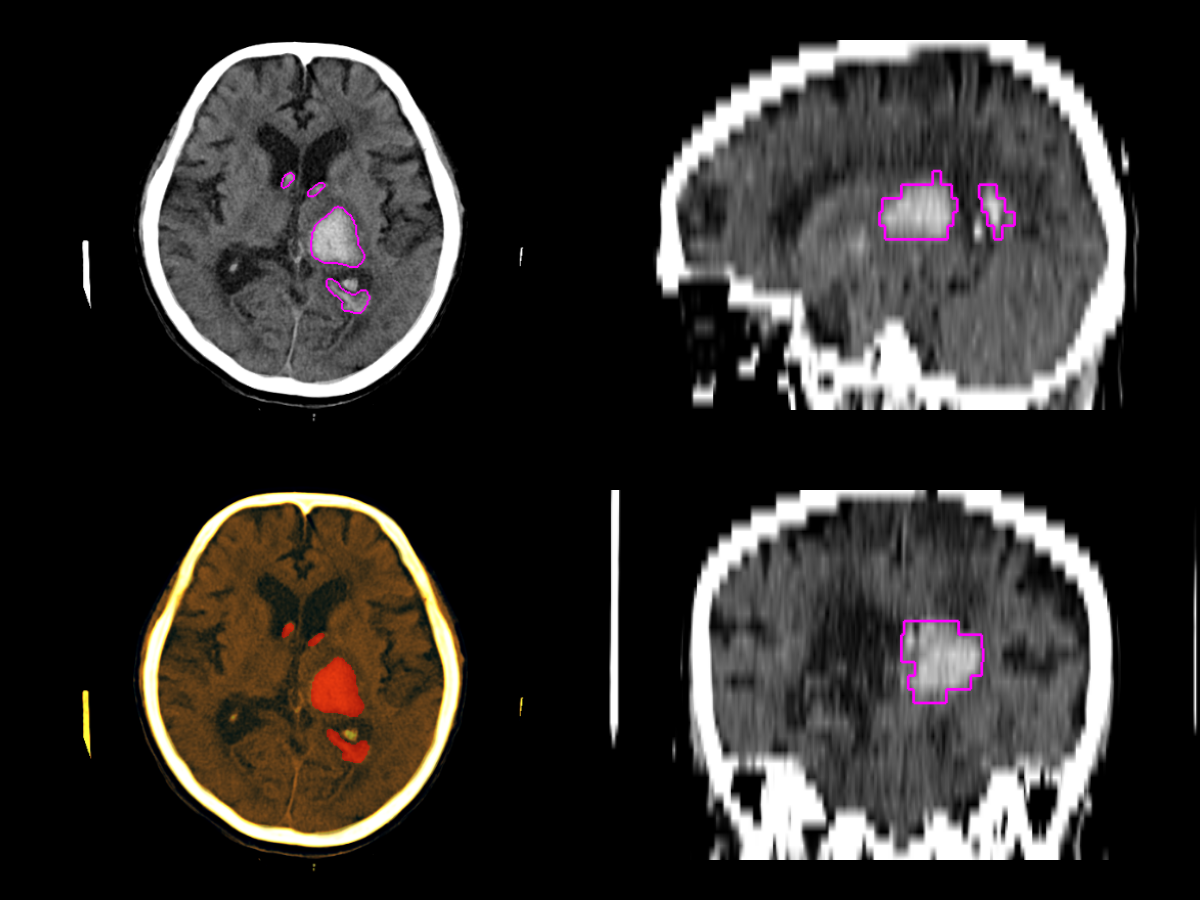}}
  \centerline{(a) DSC=0.898, HD=20.7, NSD=0.699, RVD=0.059}\medskip
\end{minipage}
\begin{minipage}[b]{0.5\textwidth}
  \centering
  \centerline{\includegraphics[width=7.5cm]{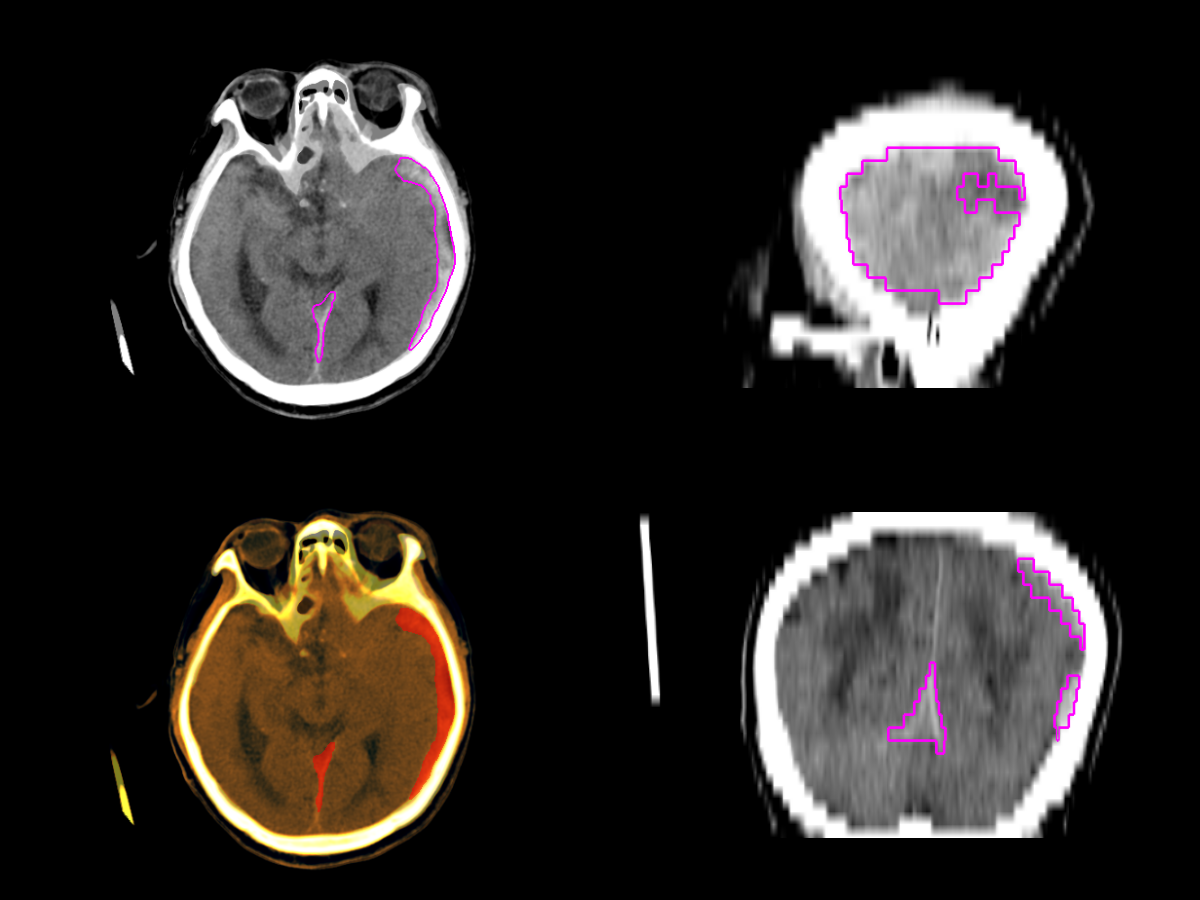}}
  \centerline{(b) DSC=0.736, HD=50.1, NSD=0.504, RVD=0.17 }\medskip
\end{minipage}
\caption{Two examples of segmented results from Viola-Unet models on online validation CT scans. We present four views of the same object: axial, sagittal, RGB-axial, and coronal. In the axial, sagittal, and coronal slices, the segmented regions were highlighted with pink edge lines, while the corresponding RGB-axial slices were marked with red color.}
\label{fig:fig4}
\end{figure*}

In table~\ref{tab3}, we show the top 10 ranking scores for INSTANCE 2022 online validation phase. Our semi-supervise trained viola-Unet-l models outperformed the comparison networks on two out of four performance metrics (i.e., NSD and RVD). An ensemble model that combined viola-Unet-l and re-implemented nnU-Net-r networks had the highest performance for DSC and HD. We show two examples of segmented results from Viola-Unet models on online validation CT scans in Fig.~\ref{fig:fig4}.
\begin{table*}
\begin{center}
\caption{Top 10 ranking scores for INSTANCE 2022 online validation phase [data extracted on 7-Aug-2022]. Note that 3 submissions provided by our team scored in the top-3. A larger version of the viola-Unet (l denotes large) was fine-tuned with self-training and achieved highest validation performance for NSD and RVD scores, while an ensemble of nnU-Net-r with viola-Unet-l was top for DSC and HD scores.} \label{tab3}
\resizebox{.8\textwidth}{!}{
\begin{tabular}{|c|c|c|c|c|}
\hline
\textbf{Models} & \textbf{DSC} $\uparrow$ & \textbf{HD} $\downarrow$& \textbf{NSD} $\uparrow$ & \textbf{RVD} $\downarrow$\\
\hline
arren & $0.7435 \pm 0.236$ & $31.616 \pm 33.221$  & $0.5201 \pm 0.153$ & $0.3580 \pm 0.450$ \\
asanner & $0.7456 \pm 0.257$ & $21.805 \pm 21.735$  & $0.5239 \pm 0.175$ & $1.1381 \pm 0.112$ \\
dongyuDylan & $0.7503 \pm 0.237$ & $29.072 \pm 26.121$  & $0.5280 \pm 0.165$ & $0.2301 \pm 0.218$ \\
testliver & $0.7537 \pm 0.236$ & $35.843 \pm 28.453$  & $0.5289 \pm 0.165$ & $0.2208 \pm 0.206$ \\
L\_Lawliet & $0.7640 \pm 0.213$ & $34.323 \pm 29.207$  & $0.5381 \pm 0.145$ & $0.2044 \pm 0.175$ \\
yangd05 & $0.7645 \pm 0.237$ & $25.725 \pm 23.801$  & $0.5403 \pm 0.169$ & $0.2322 \pm 0.235$ \\
amrn & $0.7821 \pm 0.184$ & $32.296 \pm 30.039$  & $0.5528 \pm 0.127$ & $0.2027 \pm 0.182$ \\ \hline 
nnU-Net-r (our) & $0.7943 \pm 0.174$ & $22.799 \pm 25.423$  & $0.5673 \pm 0.129$ & $0.1952 \pm 0.182$ \\
\textbf{viola-Unet-l} (our) & $0.7951 \pm 0.171$  & $24.038 \pm 29.236$  & $\textbf{0.5693} \pm 0.125$ & $\textbf{0.1941} \pm 0.179$ \\
\hdashline
Ensemble (our) & $\textbf{0.7953} \pm 0.172$  & $\textbf{21.557} \pm 25.021$  & $0.5681 \pm 0.125$ & $0.1980 \pm 0.180$ \\
\hline
\end{tabular}}
\end{center}
\end{table*}

\section{Conclusions}
\label{sec:concl}
We presented the voxels intersecting along orthogonal levels attention (Viola) module, a novel 3D attention framework that uses 3-dimensional orthogonal projections in the feature space to effectively construct fine-grained attention maps with high computational efficiency. Built upon viola module, we design a new flexible segmentation network (Viola-Unet) that can achieve high performance despite a limited training sample size in the field of biomedical imaging. On the validation dataset of the INSTANCE2022 Intracranial Hemorrhage Segmentation challenge, our Viola-Unet models outperform all other models. Importantly, the proposed network is high flexible to address different domain problems with allowing for arbitrary control of strides and kernel sizes at each layer, the number of features in both encoder and decoder layers, symmetric or asymmetric, and so on.

\section{Acknowledgments}
\label{sec:acknowledgments}
This work was supported by the foundation of Helse Sør-Øst Regional Health Authority and The Research Council of Norway under Grant 325971.

\bibliographystyle{IEEEbib}
\bibliography{strings,refs}

\end{document}